\documentclass[aps,prl,twocolumn,showpacs,showkeys,superscriptaddress]{revtex4}

\usepackage{graphicx}
\usepackage{bm}

\begin{document}
\title{Josephson junctions in narrow thin-film strips}

\author{Maayan Moshe}
\affiliation{The Raymond and Beverly Sackler School of Physics and
Astronomy, Tel Aviv University, Tel Aviv 69978, Israel}

\author{V. G. Kogan}
\affiliation{Ames Laboratory - DOE and Department of Physics and
Astronomy, Iowa state University, Ames, Iowa 50011}

\author{R. G. Mints}
\email[]{mints@post.tau.ac.il}
\affiliation{The Raymond and Beverly Sackler School of Physics and
Astronomy, Tel Aviv University, Tel Aviv 69978, Israel}

\date{\today}

\begin{abstract}
We study the field dependence of the maximum current $I_m(H)$ in narrow
edge-type thin-film Josephson junctions. The junction extends across
thin-film strip of width $W\ll\Lambda= 2\lambda^2/d$, the London depth
$\lambda\gg d$, $d$ is the film thickness. We calculate $I_m(H)$ within
nonlocal Josephson electrodynamics, which takes into account the stray
fields affecting tunneling currents. For $W\ll c\phi_0/8\pi^2\Lambda
g_c$, $g_c$ is the critical sheet current density, the phase difference
along the junction depends only on the junction geometry and the
applied field, but is independent of the $g_c$, i.e., it is universal.
Zeros of $I_m(H)$ are equidistant only in large fields (unlike the case
of junctions with bulk banks); they are spaced by $\sim\phi_0/ W^2$
that is much smaller than $\phi_0/ W\lambda$ of bulk junctions. The
maxima of $I_m(H)$ decrease as $1/\sqrt{H}$, slower than $1/H$ for the
bulk.
\end{abstract}


\par

\maketitle

The physics of the edge-type thin-film Josephson junctions (e.g., two
 films in the $(x,y)$ plane touching only along the edges at $x=0$
with no overlap) differs from that of the junctions with bulk banks
mainly because of the stray fields, that affect the currents in the
junction and in the thin-film banks. The phase difference $\varphi$
across the junction is also affected by the stray fields. As a result,
$\varphi$ is described by an integral equation, i.e., the problem
becomes {\it nonlocal}
\cite{Likharev_1,Ivanchenko_1,Mints,Ivanchenko_2,Kuzovlev_1,Kogan_1}.
\par
Development of nonlocal electrodynamics of such junctions is still in
progress and is a subject of growing interest \cite{Abdumalikov_1}.
Long-range stray-fields are relevant for physics of sequences of
interchanging $0\,$- and $\pi\,$-junctions \cite{Harlingen_1,Tsuei_1,
Hilgenkamp_1,Frolov_1,Weides_1,Mannhart_1,Smilde_1}. These anomalous
chains of tunnel Josephson junctions are studied also for the thin-film
superconductor-ferromagnet-superconductor heterostructures
\cite{Frolov_1, Weides_1}, asymmetric grain boundaries in
YBa$_2$Cu$_3$O$_{\rm 7-x}$ \cite{Mannhart_1,Hilgenkamp_1}, and
YBa$_2$Cu$_3$O$_{\rm 7-x}$/Nb zigzag junctions \cite{Smilde_1}.
\par
The phase distribution $\varphi(y)$ along thin-film edge-type junctions
has a length scale $\ell =c\phi_0/8\pi^2\Lambda g_c$, the thin-film
analog of the Josephson length \cite{Kogan_1}; $g_c$ is the critical
sheet current density, $ \Lambda =2\lambda^2/d$, $\lambda\gg d$ is the
London penetration depth, and $d$ is the film thickness. We show in
this work that when the width $W$ of the junction containing strip (and
the junction length that are the same) is less than $\ell$, the
distribution of the phase difference $\varphi(y)$ becomes $\ell$
independent, i.e., the same for junctions with different critical
currents. In other words, for $W\ll\ell$, $\varphi(y)$ is a universal
function, that depends only on the applied field and the junction
geometry.
\par
In this situation, we evaluate the field dependence of the maximum
supercurrent $I_m(H)$ through the junction that turns out quite
different from the standard Fraunhofer pattern of bulk junctions. Zeros
of $I_m(H)$ become equidistant only in large fields unlike in bulk
junctions, and are separated by $\Delta H\sim\phi_0/ W^2$, which is
much smaller than $\phi_0/ W\lambda$ of bulk junctions of the same
length. The maxima of $I_m(H)$ decrease as $1/\sqrt{H }$, that is
significantly slower than $1/H$ for the bulk. We show that $I_m(H)$ for
a SQUID made of narrow thin-film strips with edge-type Josephson
junctions differs remarkably from the canonic pattern of the bulk
junctions.
\par
Let the $x$ axis be along the strip and $z$ be perpendicular to the
film; the junction is located at $x=0$, $0\le y\le W$. The sheet
current density ${\bf g}=(g_x,g_y)$ can always be written as ${\bm
g}={\rm curl}\, S{\hat{\bm z}}=(\partial_y S\,,-\partial_x S)$, where
the $S(x,y)$ is the stream function \cite{Kogan_1}. Since the current
component normal to the edges is zero, $S$ is constant along the edges
($y=0,W$) and the total current through the strip is
\begin{equation}\label{eqn_01}
I=\int_0^W\!\!\!dy\,g_x =\int_0^W\!\!\!dy\,\partial_yS(0,y)=S(W)-S(0).
\end{equation}
\par
The London equation integrated over the film thickness reads:
\begin{equation}\label{eqn_02}
h_z + {2\pi\Lambda\over c}\,{\rm curl}_z\,{\bm g}
={\phi_0\over2\pi}\,\delta(x)\,\varphi'(y)\,,
\end{equation}
where $h_z$ consists of the applied field $H$ and the part related to
$\bm g$ by the Biot-Savart integral. The right-hand side here is a
manifestation of a general rule: the field of a Josephson junction is
equivalent to the field of a set of vortices distributed along the
junction with the line density $\varphi^\prime(y)/2\pi$
\cite{Kogan_1,Gurevich_1}.
\par
In narrow strips, the self-field due to the current $\bm g$ is of the
order $g/c$, whereas the second term on the left-hand side of
Eq.\,(\ref{eqn_02}) is of the order $g\Lambda/cW\gg g/c$. Hence, the
self-field can be disregarded, unlike the {\it applied} field $H$.
Substituting ${\rm curl}_z\,{\bm g}=-\nabla^2S$ in Eq.\,(\ref{eqn_02}),
one obtains:
\begin{equation}\label{eqn_03}
{2\pi\Lambda\over c}\,\nabla^2 S = -{\phi_0\over 2\pi}\,\delta(x)\,
\varphi'(y) + H\,.
\end{equation}
This {\it linear} equation has solutions $S=S_1+S_2$ such that
\begin{eqnarray}
\frac{2\pi\Lambda }{c\phi_0}\,\nabla^2 S_1 &=& - \frac{\delta
(x)}{2\pi} \,\varphi'(y)\,,\label{eqn_04}\\
\frac{2\pi\Lambda}{c\phi_0}\,\nabla^2 S_2&=&\frac{H}{\phi_0}\,.
\label{eqn_05}
\end{eqnarray}
The boundary condition (\ref{eqn_01}) is satisfied if $S_1(W)=S_1(0)=0$
and $S_2(W)-S_2(0)=I$. Hence we have:
\begin{eqnarray}
S_1 ({\bm r}) =\int d{\bm \rho}\,
\delta(u)\,\frac{\varphi^\prime(v)}
{2\pi}\,G({\bm r},{\bm \rho}) \,, \label{eqn_06}\\
S_2 = \frac{cH}{4\pi\Lambda}\,y(y-W) +\frac{I}{W}\,y \,.
\label{eqn_07}
\end{eqnarray}
Here, ${\bm r}=(x,y)$ and ${\bm\rho}=(u,v) $; $G({\bm r},{\bm \rho})$
is the Green's function for Eq.\,(\ref{eqn_04}) with zero boundary
conditions that satisfies $(2\pi\Lambda/c\phi_0)\,\nabla^2 G = - \delta
({\bm r}-{\bm \rho})$, an equation well studied in electrostatics
\cite{Morse_1}:
\begin{equation}\label{eqn_08}
\noindent
{\cal G} ({\bm r},{\bm \rho}) =
\tanh^{-1} \frac{\sin V \sin Y}{\cosh (X-U)-\cos Y\cos V}\,;
\end{equation}
${\cal G} = 4\pi^2\Lambda G/ c\phi_0 $, the capitals stand for
corresponding coordinates in units of $W/\pi$. The Green's function
$G({\bm r},{\bm \rho})$ gives in fact the current distribution of a
single vortex at ${\bm r}={\bm
\rho}$.
\par
Clearly, $S_1$ describes the current perturbation due to the junction.
The first term in $S_2$ represents the screening currents due to the
applied field, whereas the second is due to the field of a uniform
transport current.
\par
Given the stream function, we obtain the sheet current density through
the junction:
\begin{eqnarray}\label{eqn_09}
&&g_c\sin\varphi(y)= g_x(0,y) =\partial_y S(0,y) =\\
&&\int_0^W\!\!\!\! dv\,{\varphi'(v)\over 2\pi}\,\,\partial_y
G(0,y,0,v) +{cH\over 2\pi\Lambda}\,\left(y-{W\over 2}\right) +{I\over
W}\,.\nonumber
\end{eqnarray}
We rewrite this integral equation for the phase $\varphi(y)$ as:
\begin{equation}\label{eqn_10}
\frac{W}{ \ell} \sin\varphi =\int_0^\pi \frac{dV\,
\varphi^\prime(V) \,\sin V}{\cos Y-\cos V}
+h\left(Y-\frac{\pi}{2}\right)+ i\,,
\end{equation}
where
\begin{equation}\label{eqn_ll}
\ell = \frac{c\phi_0}{8\pi^2 \Lambda g_c}\,,\quad h =\frac{4 W^2 }
{\phi_0 }\,H\,,\quad i=\frac{8\pi^2\Lambda }{ c\phi_0}\,I
\end{equation}
are the characteristic length, the reduced field, and the
reduced current.
\par
To establish the boundary conditions for $\varphi(y)$ we employ the
London equation for $g_y(\pm 0,y)$ on the two junction banks
\begin{equation}\label{eqn_12}
g_y(\pm 0,y)= -{c\phi_0\over 4\pi^2\Lambda}\,
\left[{\partial\chi(\pm 0,y)\over \partial y}
-{2\pi\over\phi_0}A_y\right],
\end{equation}
where $\chi(x,y)$ is the phase and ${\bm A}$ is the vector potential.
We subtract these equations and utilize the continuity of $\bm A$ to
obtain $\varphi^\prime(y)\propto g_y(0,y)$. The current $g_y$ must
vanish at the junction edges, i.e.,
\begin{equation}\label{eqn_13}
\varphi^\prime(0)=\varphi^\prime(W)=0\,.
\end{equation}
\par
We note that the length $\ell$ along with $g_c$, the only material
parameter of the junction, enters only the left-hand side of
Eq.\,(\ref{eqn_10}).
\begin{figure}
\includegraphics[width=.85\columnwidth]{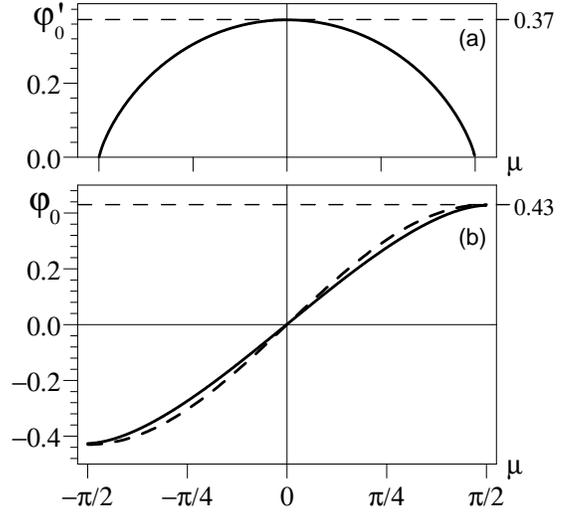}
\caption{(a) The function $\varphi'_0(\mu)$ calculated according to
Eq.\,(\ref{eqn_16}). (b) The solid line is $\varphi_0(\mu)$ obtained by
numerical integration of $\varphi'_0 (\mu)$ shown in the panel (a). The
dashed line is the approximation $\varphi_0(\mu)=0.43\,\sin\mu$.
\label{fig_1}}
\end{figure}
In narrow junctions with $W\ll\ell$, this term can be disregarded.
While neglecting the term $\propto W/\ell$ we have to disregard also
the transport current $i$; otherwise, integrating both sides of
Eq.\,(\ref{eqn_10}) over the strip does not produce identity.
\par
The truncated Eq.\,(\ref{eqn_10}) reveals a remarkable feature of
junctions in narrow strips: the phase is just proportional to the
applied field and can be written as $\varphi(y)=h\,\varphi_0(y)$ where
$\varphi_0(y)$ is an universal function governed by an integral equation
\begin{equation}\label{eqn_14}
\int_0^\pi\!\! dV\, {\varphi'_0(V)\,\sin V\over\cos Y -\cos V} + Y -
{\pi\over 2}=0,
\end{equation}
which does not contain $g_c$, the physical parameter of the
junction quality. To study this function, we introduce $s=\cos V$,
$t=\cos Y$ and write Eq.\,(\ref{eqn_14}) in the form:
\begin{eqnarray}\label{eqn_15}
\frac{1}{2\pi}\int_{-1}^1\!\! \frac{J(s)ds}{t-s}=B_n(t)\,,\\
J=2\pi\sqrt{1-s^2}\,\frac{d\varphi_0}{ds}\,,\quad B_n=- \sin^{-1}t\,.
\nonumber
\end{eqnarray}
The reason for this manipulation is this: Eq.\,(\ref{eqn_15}) is the
Biot-Savart expression for the normal component of the ``field" $B_n$
at the surface of a thin strip $-1<s<1$ carrying the ``sheet current"
$J(s)$. This integral equation can be inverted \cite{Brandt_1}. One,
however, should have in mind that the current $J(s)$ is not determined
uniquely by one field component; currents of the form $C/\sqrt{1-s^2}$
with an arbitrary constant $C$ correspond to full Meissner screening
and to zero normal component of the ``field". The latter flexibility
allows us to obtain the solution $\varphi'_0(V)$ of Eq.\,(\ref{eqn_14})
that satisfies the boundary conditions (\ref{eqn_13}):
\begin{equation}\label{eqn_16}
\varphi'_0(\mu) ={1\over \pi^2 \cos\mu}\left(2-
\int_{-\pi/2}^{\pi/2}\! \,{\eta\cos^2\eta\, d\eta\over\sin\mu
-\sin\eta} \right),
\end{equation}
where the origin is shifted to the strip middle for convenience, $\mu
=Y-\pi/2$.
\par
The integral in Eq.\,(\ref{eqn_16}) is understood as Cauchy principal
value and can be done numerically. The universal function
$\varphi'_0(\mu)$ so calculated is shown in Fig.~\ref{fig_1}\,(a). The
result of the numerical integration of this function obtained requiring
$\varphi_0(\mu)$ to be an odd function of $\mu$ is shown in
Fig.~\ref{fig_1}$\,$(b). In particular, this calculation gives
$\varphi_0(\pi/2)-\varphi_0(-\pi/2)\approx 0.86$.
\par
Thus, for any applied field in narrow thin-film junctions the phase
$\varphi(\mu)$ takes the form $\varphi(\mu)=h\varphi_0(\mu) +\theta$,
where $\theta$ is a constant. The total current through the junction is
\begin{equation}\label{eqn_17}
I=\frac{g_cW}{\pi} \int_{-\pi/2}^{\pi/2} \!\!
d\mu\,\sin[h\,\varphi_0(\mu)+\theta] \,.
\end{equation}
Maximizing this with respect to $\theta$ provides $\theta=\pi/2$ and
the maximum current $I_m$:
\begin{equation}\label{eqn_18}
{I_m\over g_cW}= \frac{1}{\pi}\Big|\int_{-\pi/2}^{\pi/2} \!\!\!\!
d\mu\,\cos[h\,\varphi_0(\mu)]\Big|\,.
\end{equation}
Hence, $I_m(H)$ can be evaluated numerically; a good approximation for
$I_m(H)$ can be obtained as follows:
\par
The odd function $\varphi_0(\mu)$ can be written as the Fourier series
$\sum a_n\sin(2n+1)\mu$ to satisfy the boundary conditions
(\ref{eqn_13}). We take the lowest approximant $\varphi_0 =a_0\sin\mu$
with $a_0=0.43$ to fit the difference $\varphi_0(W)- \varphi_0(0)=0.86$
that is found integrating numerically the exact derivative in
Eq.\,(\ref{eqn_16}). The comparison of the phase found numerically with
$a_0\sin\mu$ is shown in Fig.~\ref{fig_1}$\,$(b).
\begin{figure}
\includegraphics[width=0.85\columnwidth]{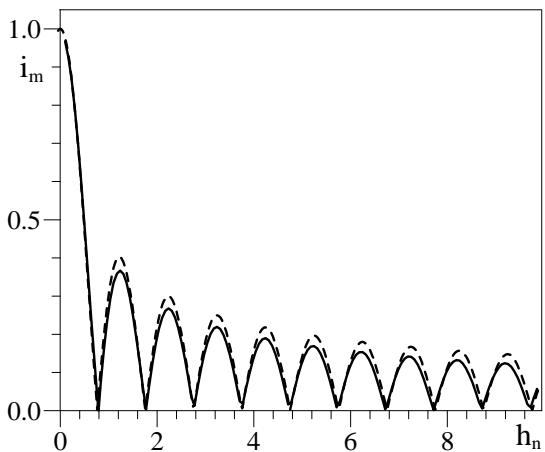}
\caption{The maximum supercurrent $i_m=I_m/g_cW$ {\it versus} the
normalized applied field $h_n=4a_0W^2H/\pi\phi_0$. The dashed line is
the approximation (\ref{eqn_19}).}
\label{fig_2}
\end{figure}
\par
In this approximation we have:
\begin{equation}\label{eqn_19}
\frac{I_m}{g_cW} ={1\over\pi}\,
\left|\int_{-\pi/2}^{ \pi/2}\!\!\!d\mu\,\cos(h\,a_0\,\sin\mu)\right|
=\left|J_0(a_0h)\right|.
\end{equation}
Figure \ref{fig_2} shows that this approximation is quite accurate as
compared to $I_m(H)$ calculated numerically with the help of
Eq.\,(\ref{eqn_18}). Zeros of the Bessel function $J_0(x)$ are
equidistant for large arguments, but they are spaced roughly by $\pi$
everywhere. Hence zeros of $I_m(h)$ are separated by $a_0\Delta
h\simeq\pi$, or in common units by:
\begin{equation}\label{eqn_20}
\Delta H\simeq 1.8\,{\phi_0\over W^2} \,.
\end{equation}
It is worth recalling that in bulk junctions of the length $W$ the
zeros are separated by $\Delta H\approx 2\phi_0/W\lambda$ that exceeds
by much the thin-film spacing.
\par
In the high-field region one can use the large argument asymptotics of
$J_0$ to obtain:
\begin{equation}\label{eqn_21}
I_m\approx 0.61\,g_c\sqrt{\phi_0\over H}\,
\left|\,\cos\left(1.72\,{ HW^2\over\phi_0} - {\pi\over 4}\right)\,\right|.
\end{equation}
Thus, the maxima of $I_m(H)$ decrease as $ 1/\sqrt{H}$, i.e., slower
than in the bulk case where $I_m\propto 1/H$.
\par
It is worth noting that in high fields the maxima $I_m(H)$ do not
depend on the junction length $W$. Qualitatively, this comes about
because the tunneling current $g_x=g_c\sin (h\varphi_0+\theta)$
oscillates fast for $h\gg 1$ so that most of the junction length does
not contribute to the total current, unlike currents in narrow belts of
the width $\delta\simeq 0.3\sqrt{\phi_0/H}$ near the strip edges.
\par
\begin{figure}
\includegraphics[width=0.85\columnwidth]{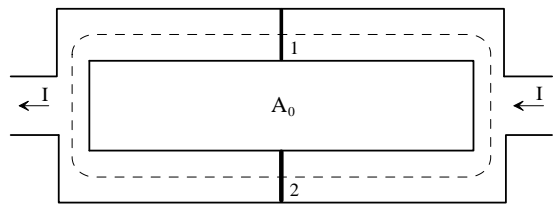}
\caption{Sketch of a rectangular SQUID made of two narrow
thin-film strips with identical edge-type junctions 1 and 2.
\label{fig_3}}
\end{figure}
\par
Let us consider now current flowing through rectangular SQUID
(superconducting quantum interference device) made of narrow thin-film
strips with two identical Josephson junctions sketched in
Fig.~\ref{fig_3}. In zero field the current distribution is symmetric
with respect to the SQUID center and the line integral of ${\bm g}$
along any symmetric contour is zero. When the field is applied, this
symmetry is violated by the screening currents. However, at the contour
in the strips middle (shown in the figure) the screening currents
vanish so that the contour integral of ${\bm g}$ remains zero. This
contour crosses the junctions at their middle, where the local
coordinates $\mu=0$. Clearly, the flux $\phi$ enclosed by this contour
does not change if the contour is shifted as a whole by $\mu$.
Integrating ${\bm g} =-(c\phi_0/4\pi^2\Lambda) (\nabla\chi +2\pi {\bm
A} /\phi_0)$ over such a contour we obtain:
\begin{equation}\label{eqn_22}
\varphi_2(\mu)-\varphi_1(\mu) = 2\pi{\phi\over\phi_0}\,.
\end{equation}
\par
The total current through the system is given by:
\begin{eqnarray}\label{eqn_23}
&& {\pi I\over g_cW}=\int_{-\pi/2}^{\pi/2} \!\!\! d\mu
(\sin\varphi_1 +\sin\varphi_2)\nonumber\\
&&=\int_{-\pi/2}^{\pi/2} \!\!\! d\mu\left[\sin(h
\varphi_0 +\theta) +\sin\left(h\varphi_0 +
\theta+\frac{2\pi\phi}{\phi_0}\right)\right] \nonumber\\
&&=2\int_{-\pi/2}^{\pi/2} \!\!\! d\mu
\sin\left(h\varphi_0 +\theta +
{\pi\phi\over\phi_0}\right)\cos\left(\frac{\pi\phi}{\phi_0}\right).
\end{eqnarray}
As above, $\theta$ is a constant with respect to which the current
should be maximized. The maximum current then corresponds to
$\theta=\pi/2-\pi\phi/\phi_0$:
\begin{equation}\label{eqn_24}
I_m=2g_cW\left|J_0\left(4a_0\,{W^2\over A_0}\,{\phi\over\phi_0}\right)
\cos\left(\pi\,{\phi\over\phi_0}\right)\right|,
\end{equation}
where $A_0$ is the area of the ``central'' contour. Note that our
argument is valid if the SQUID hole area is large relative to the area
of superconducting branches. In this case the difference between the
flux enclosed by the ``central'' contour and the SQUID hole area can be
disregarded.
\par
\begin{figure}
\includegraphics[width=0.85\columnwidth]{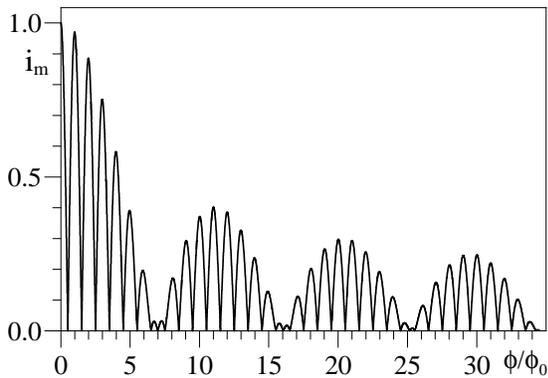}
\caption{The maximum supercurrent $i_m=I_m/2g_cW$ {\it versus} flux
$\phi/\phi_0$ for a rectangular SQUID (Fig.~\ref{fig_3}) with $A_0/W^2
=5$.
\label{fig_4}}
\par
\end{figure}

Thus, the standard SQUID pattern given by $|\cos (\pi \phi/\phi_0 )|$
is modulated in our case by a slow varying Bessel function. An example
of $I_m(\phi/\phi_0)$ is shown in Fig.\,\ref{fig_4} for a rectangular
SQUID with $A_0/W^2=5$. We stress again that the pattern shown is
obtained for large area SQUIDs made of narrow thin-film branches; for
reduced areas the interference patterns become more complex, a subject
for further study.
\par
Summarising, we have evaluated the field dependence of the maximum
supercurrent in narrow edge-type Josephson junctions in thin-film
strips; the strip width $W$ is supposed to be less than both the Pearl
length $\Lambda$ and the thin-film Josephson length $\ell$ of
Eq.\,(\ref{eqn_ll}). Calculations are done in the framework of
nonlocal Josephson electrodynamics. We demonstrate that the stray
fields cause a pattern $I_m(H)$ with much reduced distance between
zeros, $\Delta H\sim \phi_0/W^2$, and with a slow decreasing maxima in
high fields, $I_m(H)\propto 1/\sqrt{H}$. The flux dependence of the
maximum supercurrent through a SQUID made of narrow thin-film strips
with edge-type junctions differs by much from the standard periodicity.
\par
The authors are grateful to J. Mannhart and C. W. Schneider for
numerous stimulating discussions. The work of VGK at Ames Laboratory is
supported by the Office of Basic Energy Sciences of the U.S. Department
of Energy under Contract No. DE-AC02-07CH11358.
\par
\end{document}